\documentclass[%
 aip,
 amsmath,amssymb,
preprint,%
]{revtex4-1}

\usepackage{graphicx}
\usepackage{dcolumn}
\usepackage{bm}

\usepackage[utf8]{inputenc}
\usepackage[T1]{fontenc}
\usepackage{mathptmx}
\usepackage{etoolbox}

\makeatletter
\def\@email#1#2{%
 \endgroup
 \patchcmd{\titleblock@produce}
  {\frontmatter@RRAPformat}
  {\frontmatter@RRAPformat{\produce@RRAP{*#1\href{mailto:#2}{#2}}}\frontmatter@RRAPformat}
  {}{}
}%
\makeatother
\begin{document}

\preprint{AIP/123-QED}

\title[]{Probing Unknown Luttinger-Like Parameters and Band Structure in BAs: A 24-Band k·p and DFT Approach}


%

\author{Mohsen Jema\"i}
\email{mohsen.jemai@fst.utm.tn}

\author{Nouha Mastour}%

\author{Omar Zitouni}
\author{Tarek Barhoumi}

\author{Said Ridene}

\email{said.ridene@fst.rnu.tn}

\altaffiliation{Advanced Materials and Quantum Phenomena Laboratory, 
Physics Department, Sciences Faculty of Tunis, 
Tunis El Manar University, 2092 Tunis, Tunisia.}

\date{\today}

\begin{abstract}
In this work, the electronic band structure and the unknown Luttinger-like parameters associated with the p-type conduction bands (CBs) of boron arsenide (BAs) are calculated and analyzed using both 24-band $k.p$ and density functional (DFT) theories. For the DFT study, Gaussian and plane-wave basis sets implemented in the CP2K code were used. Comparison with DFT and available experimental results shows that the 24-band $k.p$ model accurately reproduces the known electronic properties of BAs. These results demonstrate that the 24-band k·p model accurately reproduces the electronic band structure and key band parameters over the investigated energy range. In addition, the 24-band $k.p$ Hamiltonian parameters are adjusted and we show that the effective masses and the band parameters are validated against previously published theoretical studies. These findings suggest that the $k.p$ method is computationally more efficient than DFT, making it suitable for modeling complex nanostructures and quantum wells based on BAs semiconductor.
\end{abstract}


\maketitle

\section{Introduction}
\label{Int_gene}

In recent years, the band structure of semiconductor materials have attracted significant interest and it has  been extensively investigated using tight-binding(TB), pseudopotential (PP), DFT, DFT+U and multiband $k.p$ methods. Indeed, accurate dispersion relations calculations within the framework of these efficient approaches were required \cite{Garcia26, Kono26,said01}. For the past two decades the multiband $k.p$ models with envelope function approximation have been successfully used by several theoretical groups in order to describe the full band structure of many interesting semiconductors \cite{Stier1999,Richard2004,Tomic2006,Kurdi2010,Tomic2010,Gladysiewicz2013,Gladysiewicz2015,Ahmed2015,Campos2018}. Furthermore, some of their parameters are not well established for certain semiconductors, particularly when considering $k.p$ models with a high band count. Therefore, comparing them with a DFT calculation is highly desirable to validate the $k.p$ method. This method is widely used in electronic device modeling, as it offers a balance between accuracy and computational cost \cite{said1,said2, said3}. Therefore, near the center of the Brillouin zone (BZ), this method is considered computationally more efficient than DFT, making it suitable for modeling complex nanostructures and devices. Obtaining the correct band structure and accurate numerical values of the band parameters values of the $k.p$ Hamiltonian can be a challenge. An accurate description of the first and second CBs across the entire BZ is of fundamental importance for correct quantitative descriptions of carrier transport phenomena \cite{Fischetti1988}, direct and indirect optical absorption calculations \cite{Ge2020}, optical and gain calculations \cite{Scharoch2021}. Moreover, some III-V semiconductor compounds such as BAs have recently attracted much interest in the field of optoelectronic devices \cite{Kudrawiec2020,Hidouri2020,Kudrawiec2019,Jaroudi2020}. Recently, the cubic structure of  BAs has been shown to overcome both limitations seen for silicon \cite{Li2022}. Specifically, it offers high electron and hole mobilities compared to silicon and has excellent thermal conductivity \cite{Lindsay2013}. The choice of BAs for the present study is explained by specific reasons such as cubic BAs was synthesized \cite{Wang2012} and its properties were also predicted from experimental and theoretical approaches \cite{Lindsay2013}. Moreover, cubic BAs exhibits very high thermal conductivity, comparable to that of diamond and graphite \cite{Lindsay2013}. In addition, among the III-V family of compounds, BAs exhibits the highest covalent character \cite{Boudjemline2011} and is stable to chemical decomposition and dissolution \cite{Osugi1966}. Moreover, another interesting feature of BAs is its valence bands (VBs) which can compensate for those of GaAs \cite{Hart2000}, making these materials ideal alloying systems. These interesting properties prompted us to further explore the electronic properties of BAs. 

Since the CBs of BAs are inverted, with the lowest CB exhibiting p-type character and the next higher band being s-type, it is necessary to determine the unknown Luttinger-like parameters that describe the p-type CBs. These parameters are essential for an accurate effective-mass and multiband $k.p$ description of the electronic structure. In particular, any theoretical investigation of BAs-based nanostructures or quantum wells relies on these parameters to model quantum confinement effects, subband dispersion, and carrier dynamics reliably. We note here, that the terms p-like and s-like refer to the orbital character of the electronic states rather than to p-type or n-type conductivity. 

In this work we apply the twenty-four $k.p$ model in order to calculate these later unknown Luttinger parameters namely $\tilde{\gamma}_{C1}$ , $\tilde{\gamma}_{C2}$ , and $\tilde{\gamma}_{C3}$ which describe the p-type CBs. Besides, for any future study of the optical transitions between either confined levels of the CB or confined levels of the VB, the so-called intersubband transitions in quantum wells, wires, dots or 2D-nanostructures based on BAs semiconductor need these important Luttinger parameters in order to apply $k.p$ multi-bands. As we have discussed in our previous work \cite{prb1} these parameters cannot be calculated by the DFT and ab-initio process because their expressions should be determined by the $k.p$ method which uses perturbation theory and Löwdin renormalization for the far bands. We have used DFT theory because the $k.p$ method requires some data which will be calculated by DFT method and to ensure that the results will be $k.p$ approach are fable. Indeed, the DFT method has been used as a companion and a support to the $k.p$ approach throughout this work. Therefore, we begin with the study of the electronic structural properties of BAs in the zinc blend phase. To study the structural properties, we used an ab-initio calculation based on DFT combined with the pseudo-potential method. The Kohn-Sham equations were solved in a self-consistent manner using a plane-wave basis implemented in the CP2K code \cite{SFWS}. To treat the exchange and correlation terms, we used the generalized gradient approximation (GGA). The obtained results are analyzed and compared with other literature results. It is important to note that in the literature especially in ref.\cite{Chimot} of N. Chimot, at al., the authors reported only the Luttinger parameters ($\gamma_1$, $\gamma_2$, and $\gamma_3$) for the VB within the 4x4 $k.p$ interaction matrix. While in our present work we have used 24 band $k.p$ Hamiltonian to cover the VB, the first CB and the second CB of BAs.

The paper is organized as follows: in section II, we present the computational approach based in the DFT method using in ab initio calculations with the CP2K code, the results and discussion of BAs band structure are detailed in section III and the conclusion is given in section IV.

\section{DFT calculation}
\begin{figure}[h]
        \centering
\includegraphics[width=0.28\textwidth]{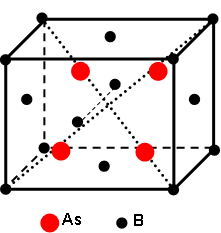}
 \caption{ Unit cell atomic structure of BAs semiconductor.}
 \label{graph_BAs}
\end{figure}

We used Gaussian and plane-wave basis sets implemented in the CP2K code which is based on the linear combination of numerical atomic orbital basis sets and pseudo-potentials to perform DFT calculations. We adopted the Perdew–Burke–Ernzerhof (PBE) exchange-correlation functional \cite{Perdew1992} because it has proven to be the most accurate for these types of semiconductors. It belongs to the class of generalized gradient approximation (GGA) functional, which is generally best suited for applications in solid-state physics. Since DFT calculations using experimental lattice constants have been reported to reproduce the band gap well of BAs \cite{Wildervanck1964, Hilli1972, Boker2001}, in our calculations we adopted this method and we have compared the calculated lattice parameters with experimental data ($a=c=4.777$ \AA). Then, we have plotted in figure \ref{graph_BAs}, the Zinc-blend crystal structure of BAs. We note that this optimization has been used for all calculations.

The electronic configurations of the constituent atoms in BAs are: ($^{33}$As : [Ar] 3d10 4s2 4p3) and Boron ($^{5}$B: [He] 2s2 2p1) which crystallizes in the cubic lattice system with a crystal parameter of $4.777$ \AA . In this structure, the conventional cell contains eight atoms occupying the following positions: B $(0, 0, 0)$, $(0, 1/2, 1/2)$, $(1/2, 0, 1/2)$, $(1/2, 1/2, 0)$ ; As $(1/4, 1/4, 1/4)$, $(1/4, 3/4, 3/4)$, $(3/4, 1/4, 3/4)$, $(3/4, 3/4, 1/4)$ in units of the lattice parameter. Electronic structure and lattice parameter calculations for BAs are performed using a $6\times 6\times 6$ Monkhorst-Pack $k$-point grid to sample the Brillouin zone. The system has been modeled using the DZVP-GTH-PBE basis set and GTH-PBE-qn pseudopotentials, both optimized for the PBE functional. To compute the Hartree and exchange-correlation contributions to the total energy and Hamiltonian, a real-space grid with a mesh cutoff energy of $500$ Ry has been applied. Structural optimization has been performed until the residual forces on each atom were below $0.01$ eV/Å, while the self-consistent field calculations have been converged to $10^{-6}$ eV. Figure \ref{BAs_Band_str_GXWKGL}, shows the band structure and the density of states (DOS) and the projected density of states (PDOS) of BAs semiconductor calculated by DFT method.

We notice that BAs has an indirect band gap of about $1.72$ eV, with the CB-minimum near the $X$-point ($\Gamma$-to-$X$ direction) and VB-maximum at the $\Gamma$-point.At this point, BAs exhibits inverted CBs, the lowest CB has a p-type character while the next higher CB has s-type character. These results are in good agreement with previous theoretical studies \cite{Buckeridge2019,Bushick2019}. Therefore, we can confirm that the CP2K code with gaussian and plane-wave basis sets reproduces appropriately the known band structure of BAs semiconductor. For the DOS and PDOS we notice that, in the upper VBs ($\approx -6$ eV to the top of VBs), the As-$4p$ and B-$2p$ orbitals are predominantly occupied. Indeed, the VB-maximum, therefore exhibits a $p$ character, mainly localized on arsenic. This is consistent with the higher electronegativity of As compared to B. In addition, the CB-minimum is dominated by B-$2p$ states, with a secondary contribution from As-$4p$ states. For the $s-$type state, we notice that is very low near the band gap, which can confirm the conduction electrons are mainly localized on boron atoms. We then determined the electron and hole effective masses of both electron and hole based on the band structure. The full set of extracted effective mass values is summarized in Table \ref{Tab5} and shows good agreement with the results reported in Ref.\cite{Bushick2019}. 

\begin{widetext}
\begin{figure}[h]
        \centering
\includegraphics[width=1.0\textwidth]{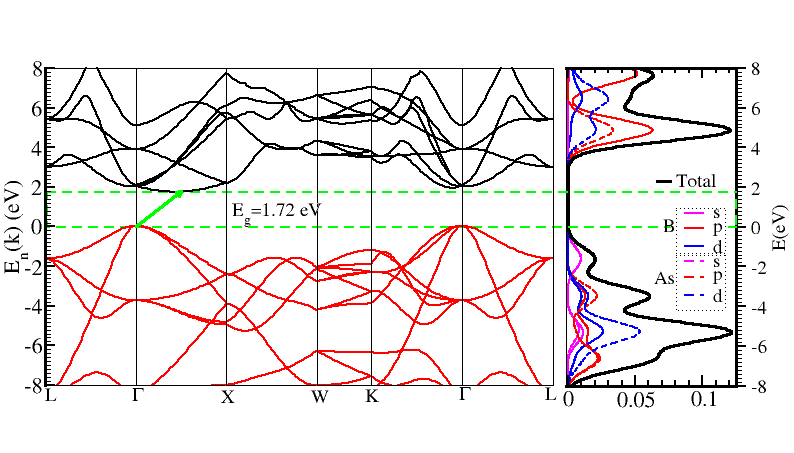}
\caption{Band structure of BAs semiconductor calculated by DFT method. The $k-$points paths are $L-\Gamma -X-W-K-\Gamma-L$ with the coordinates of the high symmetry points $X(1/2,0,0)$, $\Gamma (0,0,0)$, $L(1/2,1/2,1/2)$, $W(1/2,1/4,3/4)$, $K(3/8,3/8,3/4)$.}
\label{BAs_Band_str_GXWKGL}
\end{figure}
%
\begin{figure}[h]
        \centering
\includegraphics[width=0.6\textwidth]{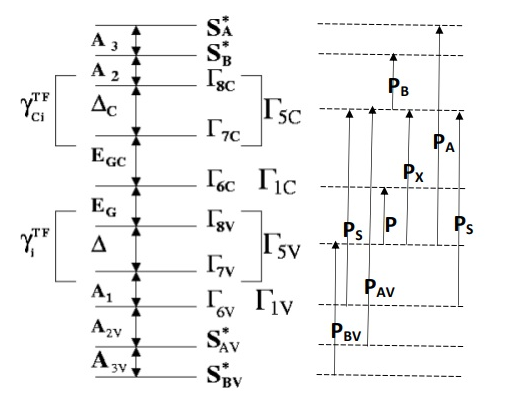}
 \caption{Schematic representation of the band parameters, Luttinger-like
parameters, and momentum matrix elements for BAs with $T_d$ symmetry. The energy
levels and their symmetry labels at the $\Gamma$ point are indicated, together
with the relevant interband coupling matrix elements. The parameters
$\gamma_i^{\mathrm{TF}}$ associated with the $\Gamma_{8}$ states and
$\gamma_i^{\mathrm{TF}}$ associated with the $\Gamma_{7}$ states denote the
Luttinger-like parameters for the valence and remote bands, respectively.
 }
 \label{Fig24model}
\end{figure}
\end{widetext}
\section{24-band $k.p$ model}
\subsection{Theoretical framework}
The purpose of this section is to compute the bulk band structure and the
associated band parameters of BAs using the 24-band ${k} \!\cdot\! {p}$
Hamiltonian. Although the formal development of this multiband model has been presented in detail in our previous works~\cite{said1,said2,said3}, we briefly recall here the essential concepts and the key physical parameters underlying the model to briefly summarize its theoretical framework and facilitate the interpretation of the results. 
Among the key adjustable quantities of the present model are the Luttinger-like parameters associated with the $p$-type CBs and $p$-type VBs, which are obtained within second-order perturbation theory. These parameters play a central role in shaping the band dispersion and are therefore crucial for an accurate description of the band structure. The Luttinger-like parameters, denoted hereafter by
$\gamma_i^{\mathrm{TF}}$ (or $\gamma_i$ for brevity), account for the coupling between the valence or CBs and energetically remote states
explicitly included in the 24-band ${k}\!\cdot\! {p}$ Hamiltonian.
Explicit analytical expressions for these parameters within the 24-band
framework have been reported in Ref.~\cite{said2,said3}. Figure \ref{Fig24model} provides a schematic illustration of the band-edge energies,
Luttinger-like parameters, and interband momentum matrix elements entering the 24-band ${k}\!\cdot\!{p}$ model of BAs.

In this figure \ref{Fig24model} we note that for $T_d$-symmetry semiconductors such as BAs, there are seven independent $k.p$ momentum matrix elements which are denoted as:
\begin{equation}
P,\; P_A,\; P_B,\; P_S,\; P_X,\; P_{AV},\; P_{BV}.
\label{eq:kp_elements}
\end{equation}
The matrix elements are:
\begin{align}
P_{AV} &= \langle A_V | p_x | C \rangle, ~~~~~~
P_{BV} = \langle B_V | p_x | C \rangle, ~~~~~~
P_{A8V} = \langle A_V | p_x | X \rangle, ~~~~~~
P_{B8V} = \langle B_V | p_x | C \rangle .
\end{align}

Recall that the matrix elements of interest in the \( H_d^{24} \) Hamiltonian are:
\begin{align}
P      &= \langle S | p_x | X \rangle, ~~~~
P_S    = \langle S_V | p_x | C \rangle, ~~~~
P_A    = \langle A | p_x | X \rangle, ~~~~
P_B    = \langle B | p_x | C \rangle, ~~~~
P_X    = \langle X | p_z | Y \rangle,
\nonumber \\
P_8    &= \langle S | p_x | C \rangle, ~~~~
P_{S8} = \langle S_V | p_x | X \rangle, ~~~~
P_{A8} = \langle A | p_x | C \rangle, ~~~~
P_{B8} = \langle B | p_x | X \rangle .
\end{align}
To each of these matrix elements we associate characteristic energy parameters,
defined by
\begin{equation}
E_\alpha = \frac{2 m_0}{\hbar^2} \, |P_\alpha|^2 ,
\label{eq:Ealpha}
\end{equation}
where $m_0$ is the free-electron mass and $P_\alpha$ represents any of the
momentum matrix elements listed in Eq.~\eqref{eq:kp_elements}.
The relations between the Luttinger-like parameters
\( \tilde{\gamma}_i \) and \( \tilde{\gamma}_{ci} \),
the momentum matrix elements,
the spin--orbit splitting energies \( \Delta \) and \( \Delta_C \),
the fundamental energy gap \( E_G \),
the intra--conduction--band energies
\( E_{GC} \), \( A_2 \), and \( A_3 \),
the intra--valence--band energies
\( A_1 \), \( A_{2V} \), and \( A_{3V} \),
and the Luttinger parameters
\( \gamma_j \) and \( \gamma_{jC} \)
are explicitly given in our previous work \cite{said1,said2,said3}.

\begin{figure}[h]
        \centering
\includegraphics[width=0.48\textwidth]
{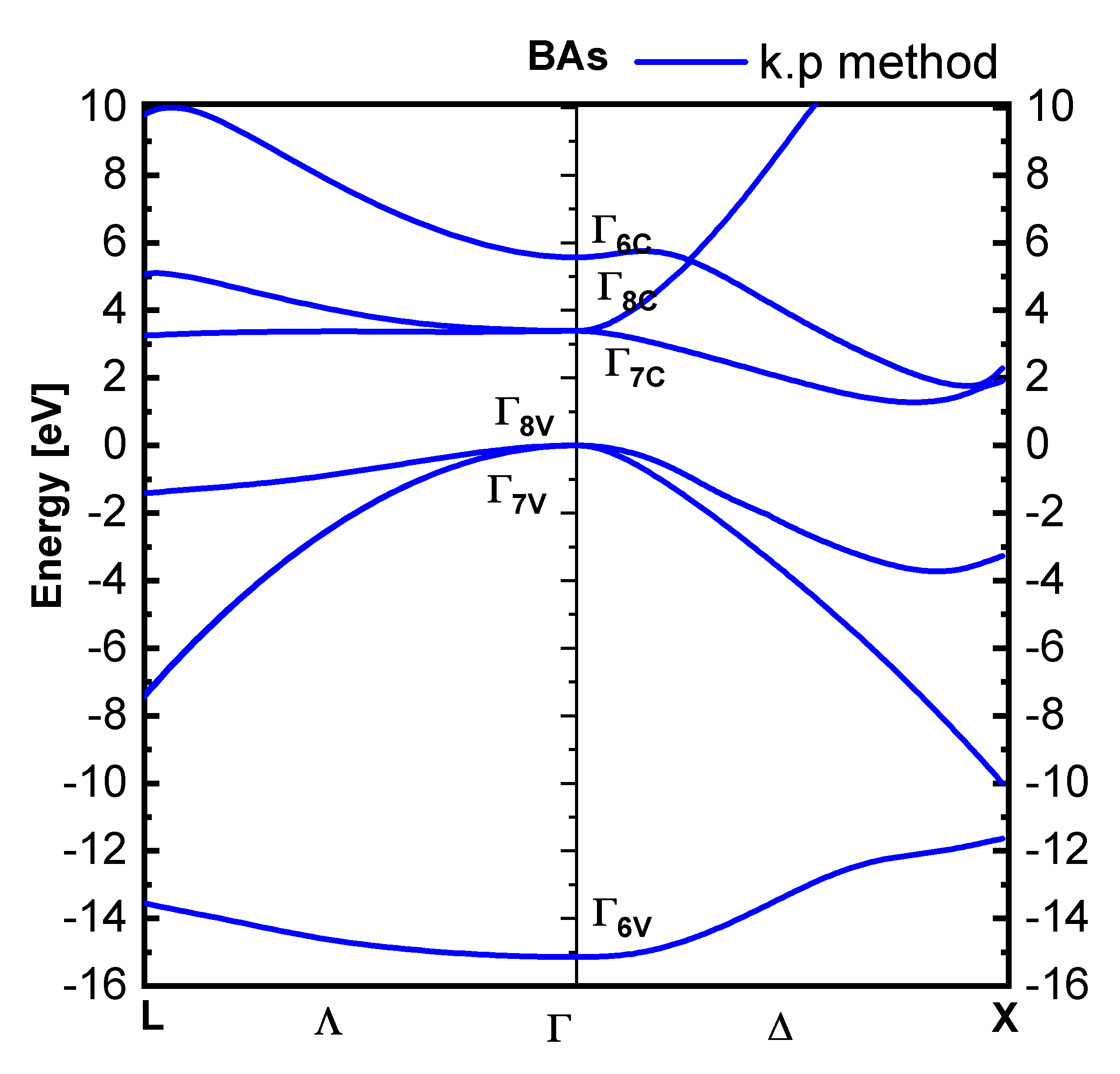}
\includegraphics[width=0.48\textwidth]
{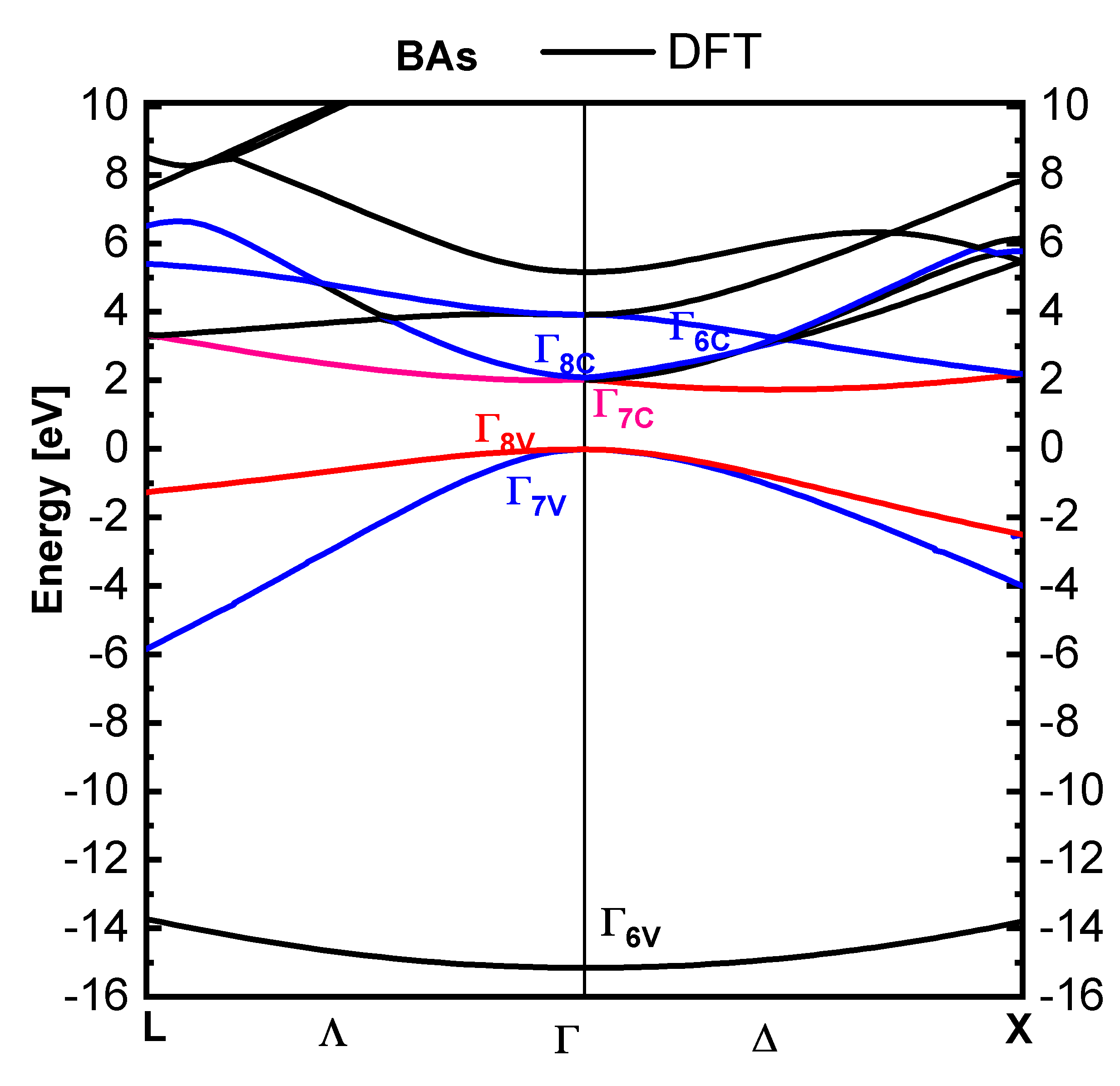}
\caption{Band structure of BAs, calculated 24-bands $k.p$ model along the $[111]$ and $[100]$ directions  of $k$-space.}
 \label{BAs55_only_kp_omar}
\end{figure}

The interband matrix elements are schematically indicated and the energies used in the text are defined as:
\begin{align}
E_G   &= E(\Gamma_6^C) - E(\Gamma_8^V), ~~
\Delta = E(\Gamma_8^V) - E(\Gamma_7^V), ~~
E_{GC} = E(\Gamma_7^C) - E(\Gamma_6^C), ~~
\Delta_C = E(\Gamma_8^C) - E(\Gamma_7^C),\nonumber \\
E_A &= E(S_A^\ast) - E(\Gamma_8^V), ~
E_B = E(S_B^\ast) - E(\Gamma_8^V), ~
E_{AV} = E(S_{AV}^\ast) - E(\Gamma_8^V), ~
E_{BV} = E(S_{BV}^\ast) - E(\Gamma_8^V).
\end{align}
However, the Luttinger-like parameters \( \gamma_i^{\mathrm{TF}} \) in the \( \Gamma_8^V \) and \( \Gamma_7^V \) levels, and \( \gamma_{Ci}^{\mathrm{TF}} \) in the \( \Gamma_8^C \) and \( \Gamma_7^C \) levels, are also indicated :
\begin{equation}
\tilde{\gamma}_1
=\tilde{\gamma}_1^{T}+\frac{1}{3}\left[\frac{E'_{PAV}}{\Delta + A_1 + A_2^{V} + A_3^{V}}+
\frac{E_{PBV}}{\Delta + A_1 + A_2^{V}}
\right]
\label{gammatilde1}
\end{equation}
\begin{equation}
\tilde{\gamma}_{\mathrm{l}}
=\tilde{\gamma}_{\mathrm{l}}^{T}+\frac{1}{6}\left[\frac{E'_{PAV}}{\Delta + A_1 + A_2^{V} + A_3^{V}}
+\frac{E_{PBV}}{\Delta + A_1 + A_2^{V}}
\right],
\qquad \mathrm{l}=2,3
\label{gammatilde2}
\end{equation}
\begin{equation}
\tilde{\gamma}_1^{C}
=\tilde{\gamma}_1^{C\,T}+\frac{1}{3}\left[\frac{E_{PAV}}
{E_G + E_{GC} + \Delta_C + \Delta + A_1 + A_2^{V} + A_3^{V}}
+\frac{E'_{PBV}}{E_G + E_{GC} + \Delta_C + \Delta + A_1 + A_2^{V}}
\right]
\label{gammatilde3}
\end{equation}
\begin{equation}
\tilde{\gamma}_{\mathrm{l}}^{C}
=\tilde{\gamma}_{\mathrm{l}}^{C\,T}+\frac{1}{6}\left[\frac{E_{PAV}}
{E_G + E_{GC} + \Delta_C + \Delta + A_1 + A_2^{V} + A_3^{V}}
+\frac{E_{PBV}}{E_G + E_{GC} + \Delta_C + \Delta + A_1 + A_2^{V}}
\right],
~\mathrm{l}=2,3
\label{gammatilde4}
\end{equation}
where the set of $\tilde{\gamma}^{T}$ (the superscript $T$
denoting ``twenty'' )parameters such as $\tilde{\gamma}_1^{T}$ and $\tilde{\gamma}_{\mathrm{l}}^{T}$($\mathrm{l}=2,3$) denote the valence-band Luttinger-like parameters,
and $\tilde{\gamma}_1^{C\,T}$ and $\tilde{\gamma}_{\mathrm{l}}^{C\,T}$ are taken from the
20-${k}\!\cdot\!{p}$ Hamiltonian which are explicitly provided in our previous Refs.~\cite{said1, said3}. 
The use of the 24-band ${k}\!\cdot\!{p}$ Hamiltonian is particularly important for the accurate calculation of the band structure and related band parameters of BAs. By explicitly including the coupling between the lowest CBs, the heavy-hole, light-hole, and split-off valence bands (VBs), as well as higher-lying remote bands, the 24-band model captures strong
interband coupling and nonparabolicity effects that are beyond the reach of
simpler few-band approximations. This comprehensive multiband description
enables a reliable determination of key band parameters, such as effective
masses, band offsets, and curvature near high-symmetry points, which are
critically sensitive to remote-band contributions. Consequently, the 24-band ${k}\!\cdot\!{p}$ Hamiltonian provides a robust and physically
consistent framework for modeling the electronic and optical properties of BAs, particularly in regimes where higher-energy bands significantly influence the low-energy dispersion. The band structures of BAs calculated within the present $24\times24\;{k\cdot p}$ model are shown in Fig. \ref{BAs55_only_kp_omar}.
The corresponding numerical values are summarized in Table~\ref{Tab1}, where selected critical-point energies are compared with those
obtained from tight-binding (TB) and pseudo-potential (PP) calculations. For both method, the band structure is accurately
reproduced over an energy range of approximately $18$~eV. In particular, the VBs are well described within a
$12$~eV window, while the two lowest CBs are reproduced within about $3$~eV along the $\Gamma$--$X$ and $\Gamma$--$L$ directions.

\begin{figure}[h]
        \centering
\includegraphics[width=0.7\textwidth]{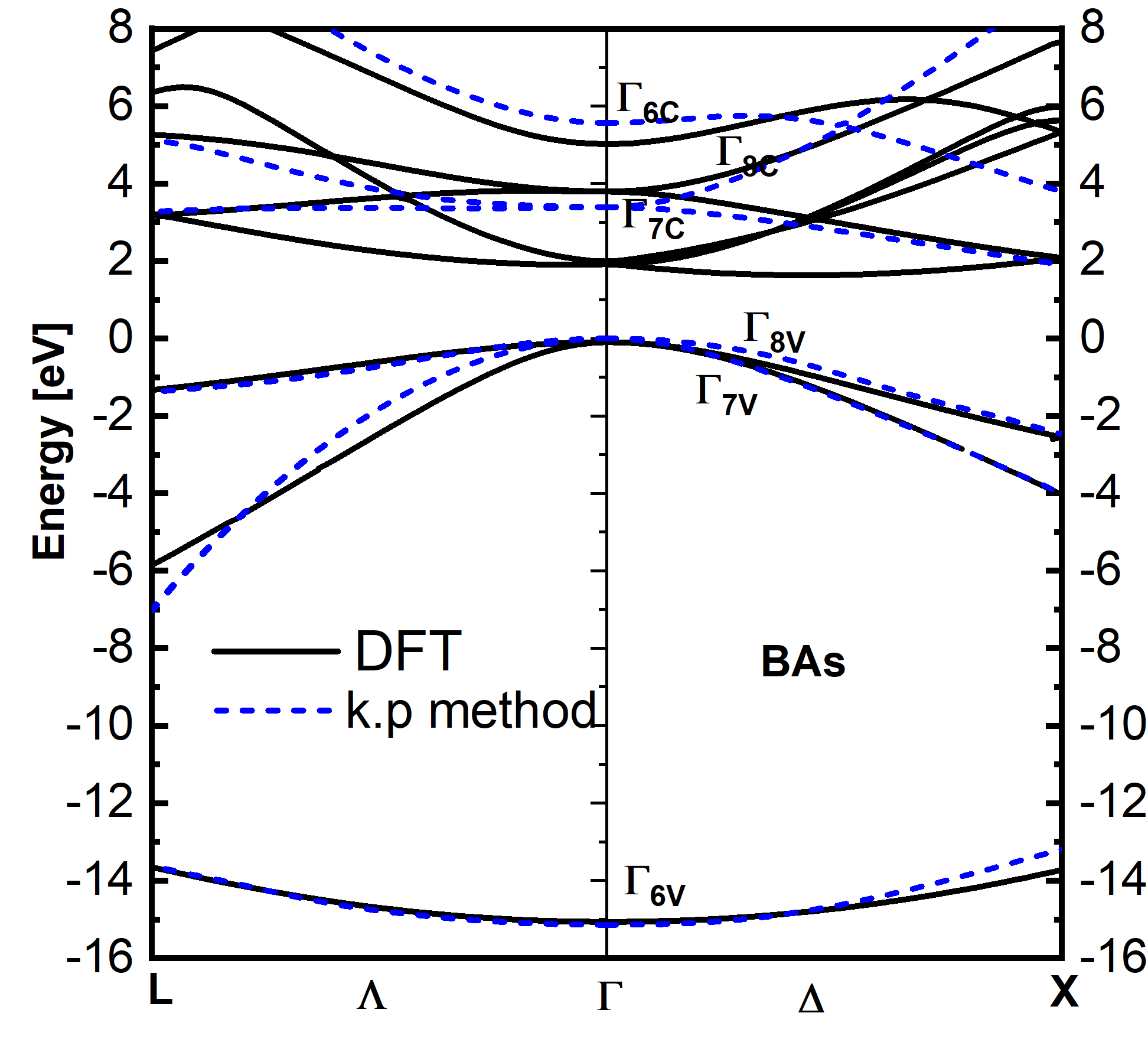}
\caption{Comparison of the band structure of BAs calculated by both $k.p$ method and DFT theory.}
 \label{BAs55_DFT_kp}
\end{figure}

In contrast to the $sp^{3}s^{*}d$ ${k\cdot p}$ model,
which provides a reliable description only within an energy
range of roughly $6$~eV, the present model yields an accurate
representation of both $s$- and $p$-like VBs as well
as the two lowest CBs. In particular, the
$X_{7V}$ and $X_{6V}$ VBs are well reproduced in the
vicinity of the $X$ point. Moreover, along the $\Gamma$--$L$
direction, the $L_{4,5V}$ and $L_{6V}$ VBs exhibit a good agreement with the corresponding TB, PP and DFT results.
Figure (FIG. \ref{BAs55_DFT_kp}) presents the band structure of BAs obtained from the
24-band ${k\cdot p}$ model, alongside the DFT results,
which were used as a reference to determine the ${k\cdot p}$
parameters. The agreement between the DFT and 24-band ${k\cdot p}$ band structures is also acceptable, particularly for the six highest VBs and the eight lowest CBs. This level of accuracy provides a sufficient
description of the electronic structure for modeling semiconductor
devices, such as LEDs, lasers, or solar cells. The corresponding \(k\cdot p\) parameters are reported in Tables~\ref{Tab1} through \ref{Tab5} below.
Generally, one can notice that these results confirm that the agreement between the DFT
and the 24-band $k.p$ band structures is moderately acceptable, particularly in the energy regions that are most relevant for device modeling (near $\Gamma$-points). For the remaining compounds, the electronic structures obtained from DFT are also well reproduced by the 24-band $k\cdot p$ model.

\subsection{Luttinger-like parameters and effective masses calculations}
The determination of electron and hole effective masses is of fundamental
importance, as these quantities govern the carrier dynamics in semiconductors.
Effective masses strongly affect carrier mobility, transport coefficients, density of states and the optical transition. Accurate knowledge of the effective masses is therefore essential
for predicting the electrical and optical performance of BAs-based devices,
including high-speed electronics and optoelectronic applications. Moreover,
effective masses provide valuable insight into the band curvature and the
strength of band coupling effects captured by the multiband ${k}\cdot {p}$
Hamiltonian, thereby serving as a sensitive benchmark for the reliability of
the underlying band-structure model.

After determining the band energies of the BAs semiconductor using the full $24\times24$ ${k}\cdot {p}$ Hamiltonian, the electron and hole effective masses were determined. The effective masses were obtained by fitting the energy dispersion near the Brillouin-zone center with a parabolic approximation. Specifically, the energies $E_n({k}=0)$ calculated from the $24\times24$ Hamiltonian were taken as the band-edge energies of the parabolic model. The dispersion relation for the $n$th electron or hole band has been assumed to be:
\begin{equation}
E_n({k}) = E_n^{24\times24}({k}=0) + \frac{\hbar^2 k^2}{2 m_n^\ast},
\end{equation}
where $m_n^\ast$ denotes the effective mass of the corresponding band and
$\hbar$ is the reduced Planck constant. The effective mass $m_n^\ast$ is treated as a fitting parameter and extracted from the curvature of the numerically calculated energy dispersion in the vicinity of ${k}=0$. It important to note that the domain of the variation of $k$ between $L$, $\Gamma$ and $X$ points of Brillouin-zone should be small enough that the effective masses considered are a good approximation of the band-edge values. This is justified by the fact that when $k$ increases in the Brillouin zone a strong nonparabolicity appears in the bands, so we limit the parabolic fitting up to about $1/4$ of the zone.
\begin{table}[h]
\caption{Parameters of BAs obtained by the $k.p$ method. Note that $E_A$, $E_B$, $E_{AV}$, $E_{BV}$, $E_P$, $E'_P$, $E_{Px}$, $E_{PS}$ represent the energies of the extrema of bands deduced from matrix elements from the Eq. (4).
\label{Tab1} }
\centering
\begin{tabular}{ccccccccc} \hline \hline 		
$E_{A}$ & $E_{B}$ & $E_{AV}$ &$E_{BV}$ & $E_{P}$ & ${E'}_{P}$ & $E_{Px}$ & $E_{PS}$ \\
8.01 & 7.90 & -16.51 & -17.01 & 14.5 & 0.55 & 12.65  & 18.502  \\
\hline
${E'}_{PS}$ & $E_{PA}$ & $E_{PB}$ &  ${E'}_{PB}$ & $E_{PAV}$ & ${E'}_{PAV}$ & $E_{PBV}$ & ${E'}_{PBV}$ \\
14.0  & 35.50 & 15.01 &  3.5 & 8.01 & 25.01 & 18.50 &5  \\
\hline
$\tilde{\gamma}_{1}$ & $\tilde{\gamma}_{2}$ & $\tilde{\gamma}_{3}$	& $\tilde{\gamma}_{C1}$ &
$\tilde{\gamma}_{C2}$ & $\tilde{\gamma}_{C3}$ & $\tilde{\gamma}_{C}$ &$\tilde{\gamma}_{V}$ \\
-0.240 &-0.545 &-0.62 &-1.25 &0.22 &-0.405 & -0.55 & 2.380 \\
\hline
$\alpha_{A}$ & ${\alpha '}_{A}$ &${\alpha ''}_{A}$ & $\beta_{A}$ & $\alpha_{AV}$ & 
${\alpha '}_{AV}$ & ${\alpha ''}_{A}$ & $\beta_{AV}$  \\
-3.95 & 3.841 & 0.44 & 3.9 & -1.55 & -5.5 & 0.10 & 4.16  \\
\hline
$\alpha_{B}$ & ${\alpha '}_{B}$ & ${\alpha ''}_{B}$ & $\beta_{B}$ & $\alpha_{BV}$ & ${\alpha '}_{BV}$ & ${\alpha ''}_{BV}$ &$\beta_{ABV}$ \\
-0.398 & 2.651 & -0.06 &  0.02 & -2.01 & -1.2 &-2.00 &0.21   \\
\hline \hline
\end{tabular}
\end{table}	
Thus, we obtain the following values for the Luttinger parameters: $\gamma_1=4.755$, $\gamma_2=0.087$ and $\gamma_3=1.403$.
We now turn to the Luttinger parameters of the $\Gamma_8^V$ VB ($\gamma_i$) and the 
Luttinger-like parameters of the $\Gamma_8^C$ CB ($\gamma_{iC}$). 
The $\gamma_i$ values are determined using a set of 41 adjustable parameters through expressions of the Luttinger parameters given in Eqs. {\ref{gammatilde1}, \ref{gammatilde2}, \ref{gammatilde3} and \ref{gammatilde4}). 
For BAs, these formulas yield $\gamma_1 = 4.38$, $\gamma_2 = 0.08$, and $\gamma_3 = 1.27$, in good agreement with the results reported in the Ref.\cite{Bushick2019, Buckeridge2019, Acharya2023}.  
These findings demonstrate that the Luttinger parameters derived in this work 
are fully compatible with previously published calculations~\cite{Wang2012,Boudjemline2011}.

\begin{table*}[h!]
\caption{Comparison of BAs band gap calculated with different methods. 
\label{Tab2} }
\centering
\begin{tabular}{ccc} 
\hline \hline 
Method & Minimum direct band gap & Indirect band gap
\\ 
$sp^{3}s^{*}d$ ${k\cdot p}$ (this work)  & 3.390  & 1.278 \\
LDA\cite{Bushick2019,Buckeridge2019} & 3.328 & 1.189  \\
LDA+SO \cite{Bushick2019,Buckeridge2019}& 3.128 & 1.124  \\
GW+SO \cite{Bushick2019,Buckeridge2019} & 4.135  & 2.049  \\
HSE06 \cite{Buckeridge2019}& 3.846 & 1.693 \\
GGA-PBE+D3 (this work) & 3.870 & 1.720  \\
Experiment \cite{Lyons2018} & - & 1.77 \\
\hline \hline
\end{tabular}
\end{table*}
Consequently, we have quantified the agreement by comparing the bandgap energies. Indeed, the calculated bandgap obtained from our DFT(GGA+PBE+D3) method is $1.720$ eV, DFT(HSE06) method is $1.693$ eV and 24-band $k·p$ model is $1.278$ eV, while the corresponding value predicted by experimental one is 1.77 eV, resulting in a relative deviation of only $1.6$ \% with HSE06 and $\sim 3$\% compared to the experimental value. Therefore, for the bandgap our results are in good agreement with the experimental one with the exception of $k.p$, method which is slightly different. The quantitative comparison between the DFT and 24-band $k.p$ are in good agreement, therefore, the relative deviation of $1.5$\% and $12.4$ \% for the energies of the extrema of bands values and effective masses.
\begin{table}[h!]
\caption{Comparison of the energies obtained with different methods calculations, on the $X$ and $L$ points. 
\label{Tab3} }
\centering
\begin{tabular}{cccccc} 
\hline \hline 
Energy (eV) & $sp^{3}s^{*}d$ ${k\cdot p}$ & LDA\cite{Bushick2019,Buckeridge2019}  & GW\cite{Bushick2019, Buckeridge2019} &
GW +SO\cite{Bushick2019, Buckeridge2019} & GGA-PBE+D3  \\ 
 &(this work) & &  &  &  (this work) \\ 
$X_{6V,1}$ & -11.820  &-11.192 &-11.384 &-11.384 & -10.468 \\
$X_{7V}$   & -8.793  &-8.374  & -8.806 &-8.808  & -8.015 \\
$X_{6V,2}$ & -3.828  &-4.029  & -4.205 & -4.275 & -3.922 \\
$X_{6C}$   &  1.966  &  1.419 & 2.264  & 2.264  &  2.166 \\ 
$X_{7C}$   &  1.978  & 1.444  & 2.356  & 2.356  & 2.166  \\
$L_{6V,1}$ & -13.580 &-12.469 &-12.688 &-12.688 &-12.039 \\
$L_{6V,2}$ & -7.230  &-8.696  &-9.002  & -9.003 & -8.438 \\ 
$L_{6V,3}$ & -1.563  &-1.788  & -1.910 & -1.985 &-1.642\\
$L_{6C,1}$ & 3.258  & 2.841  & 3.658  & 3.658  & 2.979 \\
$L_{6C,2}$ & 5.099   & 4.811  &  5.826 & 5.789  & 5.372 \\
\hline \hline
\end{tabular}
\end{table}

\begin{table}[h!]
\caption{Comparison of the energies obtained using the 24- bands ${ k.p}$ method and extracted directly from the DFT calculations, in the $\Gamma$ point. 
\label{Tab4} }
\centering
\begin{tabular}{ccccccc} \hline \hline 
Energy (eV) & $E_{\Gamma 8C}$ & $E_{\Gamma 7C}$ & $E_{\Gamma 6C}$	&$E_{\Gamma 8V}$	&
$E_{\Gamma 7V}$	& $E_{\Gamma 6V}$ \\
$sp^{3}s^{*}d$ ${k\cdot p}$ & 3.39 & 3.39 & 5.56 & 0.00 & 0.00 & -15.14 \\
GGA-PBE+D3   & 3.87  & 3.87  & 5.01  & 0.00 & 0.00  & -14.91 \\
$\%$ diff. & 12.4 & 12.4 & 9.9 & 0 & 0 & 1.5\\ 
\hline \hline
\end{tabular}
\end{table}

\begin{table}[h!]
\caption{Comparison of the effective mass for different points of high symmetry obtained using the 24- bands $k.p$ method and extracted directly from the DFT calculations of BAs. 
\label{Tab5} }
\centering
\begin{tabular}{ccccccc} \hline \hline
 $m^*/m_0$ & ${\Gamma 8C}$ & ${\Gamma 7C}$ & ${\Gamma 6C}$	&${\Gamma 8V}$	&
${\Gamma 7V}$	& ${\Gamma 6V}$ \\
$sp^{3}s^{*}d$ ${k\cdot p}$ & 0.4477 & 0.1873 & 0.5456 & 0.5907 & 0.2207 & 2.0571  \\
GGA-PBE+D3 & 0.3403 & 0.2968 & 0.4081 & 0.6421 & 0.2457 & 1.2210 \\
\hline \hline
\end{tabular}
\end{table}
\newpage
\section{Conclusion}
We have used both $24\times 24$ $k.p$ Hamiltonian and DFT to calculate BAs band structure. We obtained the corresponding parameters using both methods. Using this $k\cdot p$ model, we achieved a quantitatively accurate description of all VBs as well as the two lowest CBs, both in terms of their energies and band curvatures. Our results show that the Luttinger parameters are well captured by this $k.p$ approach. We demonstrated very good agreement between the DFT and the $k.p$  results. We have also determined the effective masses from the band structure. Finally, we have also compared the parameters obtained from the fitting and calculated directly from the DFT (GGA-PBE+D3) momentum matrix elements.We found good agreement across most of these parameters; specifically, our calculations using the GGA-PBE+D3 method yielded a band gap energy remarkably close to the HSE06 result, with a percentage difference of only 1.6\%, and within $\sim 3$ of the experimental value.

\section*{Acknowledgments} We would like to thank Prof.Habib Bouchriha from University of Tunis El Manar for his useful discussions.
\section*{Declarations}
Data Availability Statement: No data are associated with this manuscript.
\section*{Conflict of interest}
The authors declare that they have no known competing financial interests or personal relationships that could have appeared to influence the work reported in this paper.
\section*{Author Contributions}
All authors contributed to the study conception and design. Material preparation, data collection and analysis were performed by all authors. The first draft of the manuscript was written by Said Ridene, and all authors commented on previous versions of the manuscript. All authors read and approved the final manuscript.

\end{document}